\documentclass[prb,preprint,superscriptaddress,nobibnotes]{revtex4}
\usepackage{graphicx}% Include figure files
\usepackage{dcolumn}% Align table columns on decimal point
\usepackage{bm}% bold math
\usepackage{times,mathptmx}
\usepackage{color}
\usepackage{multirow}
\usepackage{enumitem}
\usepackage{chngpage}
\begin{document}

%\preprint{APS/123-QED}

\title{\bf High-quality lithium niobate photonic crystal nanocavities}% Force line breaks with \\

\author{Hanxiao Liang}
\thanks{These authors contributed equally to this work.}
\affiliation{Department of Electrical and Computer Engineering, University of Rochester, Rochester, NY 14627}
\author{Rui Luo}
\thanks{These authors contributed equally to this work.}
\affiliation{Institute of Optics, University of Rochester, Rochester, NY 14627}
\author{Yang He}
\affiliation{Department of Electrical and Computer Engineering, University of Rochester, Rochester, NY 14627}
\author{Haowei Jiang}
\affiliation{Department of Electrical and Computer Engineering, University of Rochester, Rochester, NY 14627}
\affiliation{School of Physics and Astronomy, Shanghai Jiao Tong University, Shanghai, China 200240}
\author{Qiang Lin}
\email{qiang.lin@rochester.edu}
\affiliation{Institute of Optics, University of Rochester, Rochester, NY 14627}
\affiliation{Department of Electrical and Computer Engineering, University of Rochester, Rochester, NY 14627}

%\date{}% It is always \today, today,


\begin{abstract}
Lithium niobate (LN) exhibits unique material characteristics that have found many important applications. Scaling LN devices down to a nanoscopic scale can dramatically enhance light-matter interaction that would enable nonlinear and quantum photonic functionalities beyond the reach of conventional means. However, developing LN-based nanophotonic devices turns out to be nontrivial. Although significant efforts have been devoted in recent years, LN photonic crystal structures developed to date exhibit fairly low quality. Here we demonstrate LN photonic crystal nanobeam resonators with optical Q as high as 10$^5$, more than two orders of magnitude higher than other LN nanocavities reported to date. The high optical quality together with tight mode confinement leads to extremely strong nonlinear photorefractive effect, with a resonance tuning rate of $\sim$0.64~GHz/aJ, or equivalently $\sim$84~MHz/photon, three orders of magnitude greater than other LN resonators. In particular, we observed intriguing quenching of photorefraction that has never been reported before. The devices also exhibit strong optomechanical coupling with gigahertz nanomechanical mode with a significant ${\rm f \cdot Q}$ product of $1.47\times 10^{12}$~Hz. The demonstration of high-Q LN photonic crystal nanoresonators paves a crucial step towards LN nanophotonics that could integrate the outstanding material properties with versatile nanoscale device engineering for diverse intriguing functionalities.

\end{abstract}

\maketitle

Lithium niobate (LN) exhibits outstanding electro-optic, nonlinear optical, acousto-optic, piezoelectric, photorefractive, pyroelectric, and photoconductive properties \cite{Gaylord85} that have found very broad applications in telecommunication \cite{Wooten00}, nonlinear/quantum photonics \cite{Fejer95, Gisin07}, microelectromechanics \cite{Pijolat09, Piazza13}, information storage \cite{Hesselink94, Psaltis98}, sensing \cite{Reindl04}, among many others \cite{Arizmendi04}. Recently, significant interest has been attracted to develop LN photonic devices on chip-scale platforms \cite{Gunter07, Gunter12, Niu12, Fathpour13, Reano14, Fathpour14, Loncar14, Cheng15, Pertsch15, Xu15, Hu15, Yang15, Shayan16, Jiang16, Bower16, Bower17, Fathpour17, Loncar17, Luo17, Amir17}, which have shown significant advantage in device engineering compared with conventional approaches. Miniaturization of device dimensions dramatically enhances optical field in the devices which enables a variety of nonlinear optical, quantum optical, and optomechanical functionalities. %\cite{Jiang16, Bower16, Fathpour17, Loncar17, Silberhorn13, Zhu14}.

Among various approaches developed to date, photonic crystal is probably one of the most efficient ones for light confinement \cite{PhCBook}, which has been demonstrated on a variety of material platforms \cite{PhCBook, Noda07, Lalanne08, Notomi09}. For lithium niobate, however, it remains an open challenge to achieve high optical quality, primarily due to significant challenge in device fabrication \cite{Baida05, Gu06, Salut06, Gunter09, Pertsch10, Laude10, Baida11, Bernal12, Diziain13, Wang14, Pertsch14}. The LN photonic crystal nanocavities demonstrated to date generally exhibit low optical Q in the order of $\sim 100$ \cite{Bernal12, Diziain13, Pertsch14}, which seriously limits their potential applications.

An alternative approach to get around the fabrication challenge is to fabricate waveguide structures on a different material deposited on top of a LN substrate to provide wave guidance while using LN as a cladding material \cite{Fathpour13, Reano14, Fathpour14, Hu15, Yang15, Shayan16, Bower16, Bower17, Fathpour17, Amir17}. This approach, however, limits the extent of optical mode overlap with the LN layer as well as the design flexibility of waveguide structure, due to the limitation of index contrast required between the waveguide material and the LN substrate.

\begin{figure}[hbtp]
	\includegraphics[width=1\columnwidth]{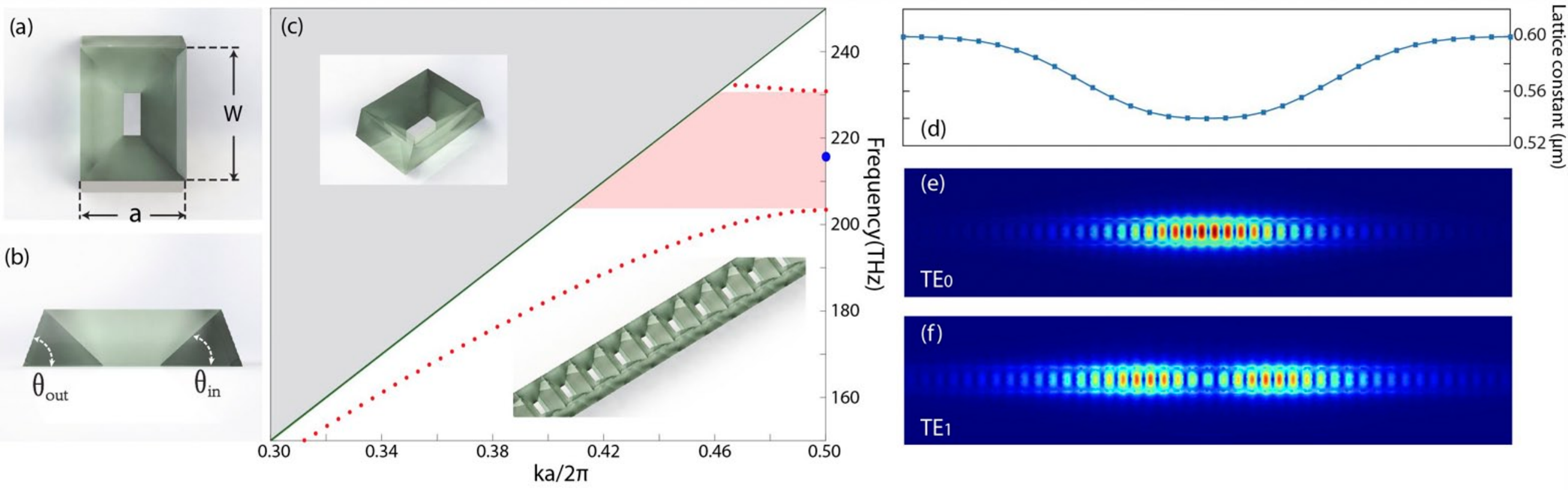}
	\caption{\label{Fig1} Properties of the photonic band structure and defect cavity modes of the designed LN photonic crystal nanobeam. \textbf{a.} Top view of the schematic of the unit cell of the photonic crystal nanobeam. {\bf b.} Cross section of the unit cell, showing the inside and outside sidewall angles. {\bf c.} Band structure of the designed photonic crystal nanobeam (red dotted curves). The green solid line corresponds to the light line. The pink region indicates the photonic bandgap and the blue dot indicates the resonance frequency of the fundamental defect cavity mode. The inset on the lower right corner shows the 3-D schematic of the photonic crystal nanobeam and that on the upper left corner shows that of the unit cell. {\bf d.} Lattice constant as a function of position, which is optimized for high radiation-limited optical Q. {\bf e and f.} The optical mode field profiles of the fundamental (TE0) and second-order (TE1) transverse-electric-like (TE-like) cavity modes, with electric field dominantly lying in the device plane. The mode field profiles are simulated by the finite element method. Note that the horizontal axes of (e) and (f) have a different scale from that of (d).}
\end{figure}

In this paper, we demonstrate LN photonic crystal nanobeam resonators with optical Q up to $1.09\times 10^5$, more than two orders of magnitude higher than any other LN photonic crystal nanocavities reported to date \cite{Baida05, Gu06, Salut06, Gunter09, Pertsch10, Laude10, Baida11, Bernal12, Diziain13, Wang14, Pertsch14}. The high optical Q together with the tiny effective mode volume ($\sim 1.03(\lambda/n)^3$) leads to extremely strong nonlinear photorefractive effect, with a resonance tuning rate of $\sim$0.64~GHz/aJ, corresponding to $\sim$84~MHz/photon, three orders of magnitude greater than other LN resonators \cite{Maleki06, Breunig16}. In particular, it enables us to observe the intriguing quenching of photorefraction that has never been reported before. It also results in strong coupling between the optical cavity mode and the mechanical motion of the device structure, which allows us to sensitively probe the rich nanomechanical properties of the LN photonic crystal nanobeams up to $\sim$1~GHz. The demonstration of high-Q LN photonic crystal nanocavities paves the foundation towards LN nanophotonics that would combine elegantly the unique material properties of lithium niobate and versatile nanophotonic device design/fabrication, for broad nonlinear photonic, quantum photonic, optoelectronic, and optomechanical applications.

\section*{Device Design and Fabrication}

\begin{figure}[btp]
	\includegraphics[width=0.5\columnwidth]{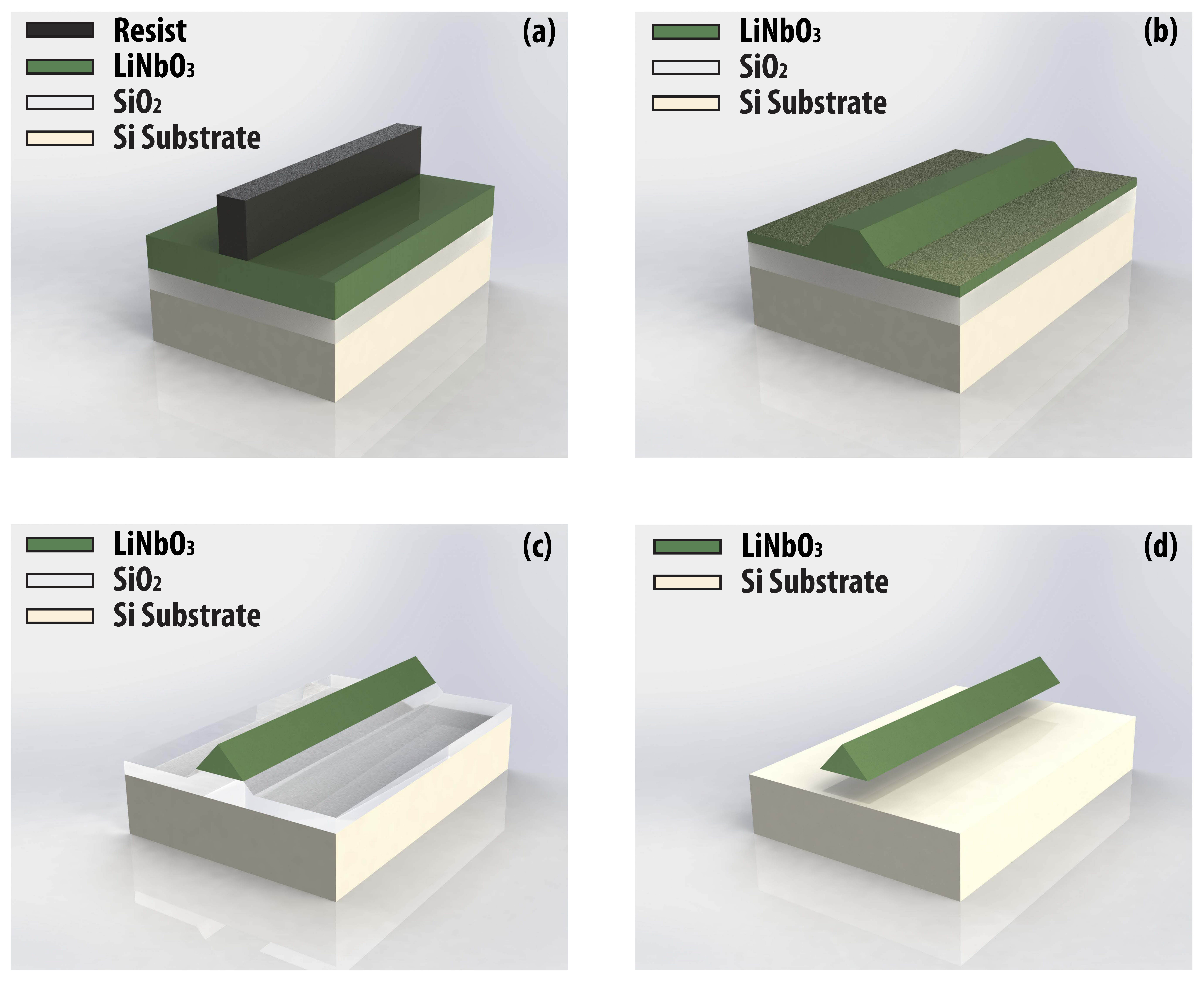}
	\caption{\label{Fig2} Over-etching process to produce desired device structure. \textbf{a.} Structure patterning on the ZEP mask by electron-beam lithography. {\bf b.} Ar-ion milling to produce trapezoid-shaped cross section. {\bf c.} Further Ar-ion milling to thin down the device layer thickness to form triangular shaped cross section. {\bf d.} undercutting of buried oxide layer by diluted hydrofluoric acid.}
\end{figure}

\begin{figure}[btp]
	\includegraphics[width=1.0\columnwidth]{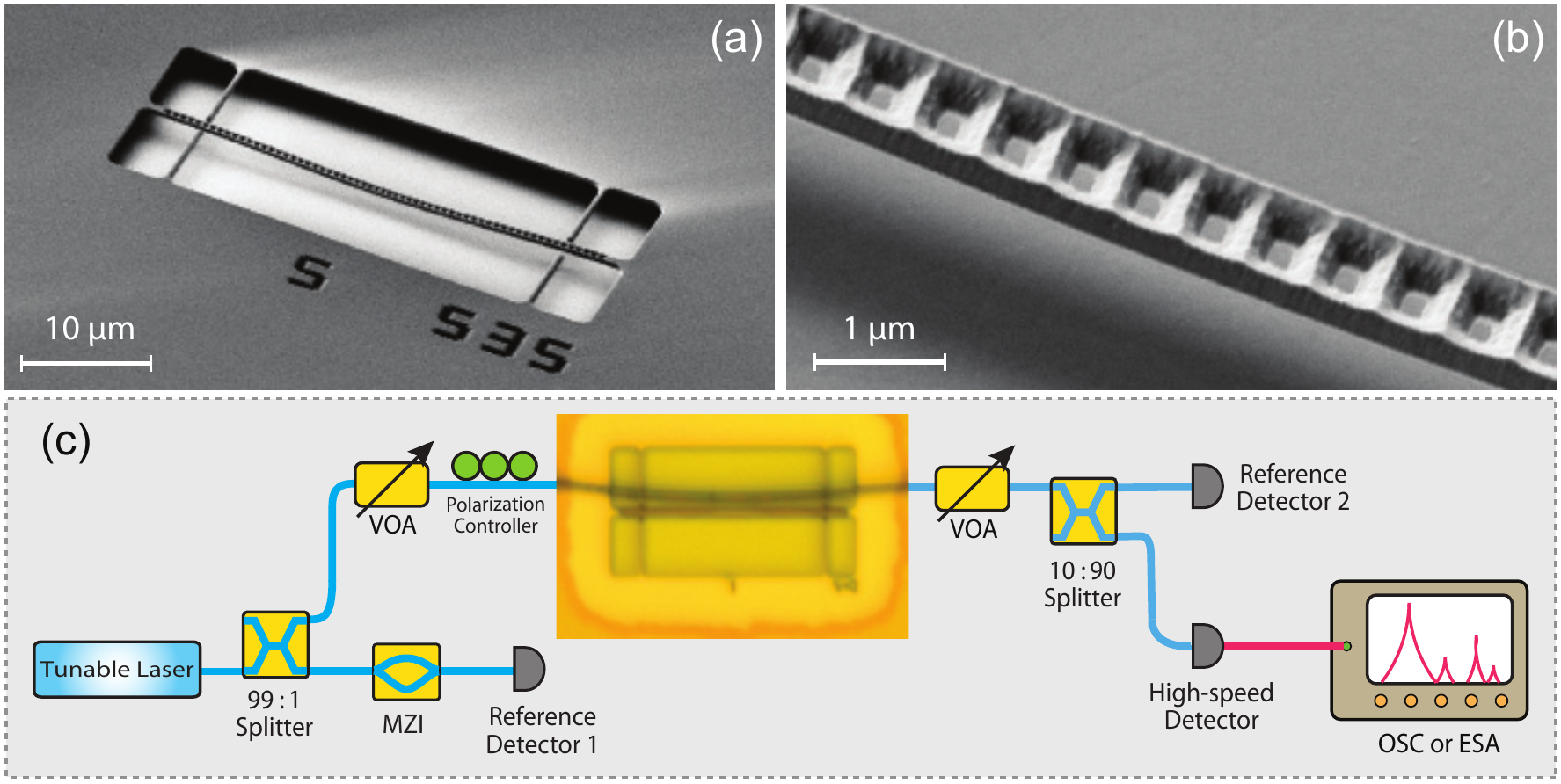}
	\caption{\label{Fig3} Fabricated device structure and experimental testing setup. {\bf a.} Scanning electron microscopic image of a fabricated LN photonic crystal nanobeam. {\bf b.} Zoom-in image of a section of the photonic crystal nanobeam. {\bf c.} Schematic of the experimental testing setup. MZI: Mach-Zehnder interferometer, used to calibrate the laser wavelength; VOA: variable optical attenuator; OSC: oscilloscope; ESA: electrical spectrum analyzer. The inset shows an optical microscopic image of a device coupled to a tapered optical fiber that is mechanically supported by two nanoforks fabricated nearby. }
\end{figure}

Current plasma etching approaches to fabricate high-quality LN photonic devices generally produce a slant angle on the device sidewall \cite{Loncar14, Jiang16}. Although it might help improve the optical quality of LN microresonators, it impacts seriously on LN photonic crystals which have stringent requirement on the precision of device fine structures. To achieve high optical Q, we tailored our design to incorporate such slant angle into the structure of photonic crystals. The insets of Fig.~\ref{Fig1}(a) show the rectangular-shaped unit cell of the designed photonic crystal nanobeam (Fig.~\ref{Fig1}(c), inset), where the angles of inside and outside sidewalls (Fig.~\ref{Fig1}(b)), $\theta_{in}$ = 45$^\circ$ and $\theta_{out}$ =75$^\circ$, are determined by the plasma etching process. The width $W$ of the nanobeam, the layer thickness $H$, and the lattice constant $a$ are the free parameters which we optimized to produce an optimal bandgap. Figure \ref{Fig1}(c) shows the band diagram simulated by the finite element method, where a LN photonic crystal nanobeam with dimensions of $W=750$~nm, $H=250$~nm, and a lattice constant of $a=600$~nm exhibits a bandgap of 28 THz covering optical frequency from 203 to 231 THz, for the transverse-electric-like (TE-like) polarization with the electric field dominantly lying in the device plane.

To produce a defect cavity, we gradually decreased the lattice constant from 600~nm to 540~nm around the center of the nanobeam. We optimized the nanobeam with a pattern of lattice constants as shown in Fig.~\ref{Fig1}(d), which results in a localized defect cavity at the center of the nanobeam whose fundamental cavity mode exhibits a resonance frequency close to the center of the photonic bandgap, as indicated by the blue dot in Fig.~\ref{Fig2}(c). Figure \ref{Fig1}(e) and (f) show the optical mode field profiles of the fundamental (TE0) and second-order (TE1) TE-like cavity modes, simulated by the finite element method. The simulations show that the two modes exhibit radiation-limited optical Qs of $6.0\times 10^6$ and $5.2\times 10^5$, respectively, with effective mode volumes as small as 1.03$(\lambda/n)^3$ and 1.80$(\lambda/n)^3$ (where $\lambda$ is the optical resonance wavelength and $n$ is the refractive index).

Our devices were fabricated on a 300-nm-thick x-cut congruent single-crystalline LN thin film sitting on a 2-${\mu m}$-thick buried oxide layer. The structure was patterned with ZEP-520A positive resist as a mask via electron beam lithography (Fig.~\ref{Fig2}(a)) and was etched with the Ar-ion milling process \cite{Loncar14, Jiang16}. We developed an over-etching process to produce desired fine structures and sidewall smoothness, as schematically shown in Fig.~\ref{Fig2}(b)-(d). During the beginning stage of etching, the Ar-ion milling process produces slant angles on the device sidewall, leading to a trapezoid-shaped cross section (Fig.~\ref{Fig2}(b)). Further Ar-ion milling etched the ZEP-520A mask away and thinned the thickness of the LN layer down to $\sim$250~nm, eventually forming a triangularly-shaped cross section (Fig.~\ref{Fig2}(c)). Finally, the buried oxide layer was undercut by diluted hydrofluoric acid to form a suspended photonic crystal nanobeam (Fig.~\ref{Fig2}(d)).

\section*{Linear optical properties}

\begin{figure}[btp]
	\includegraphics[width=0.8\columnwidth]{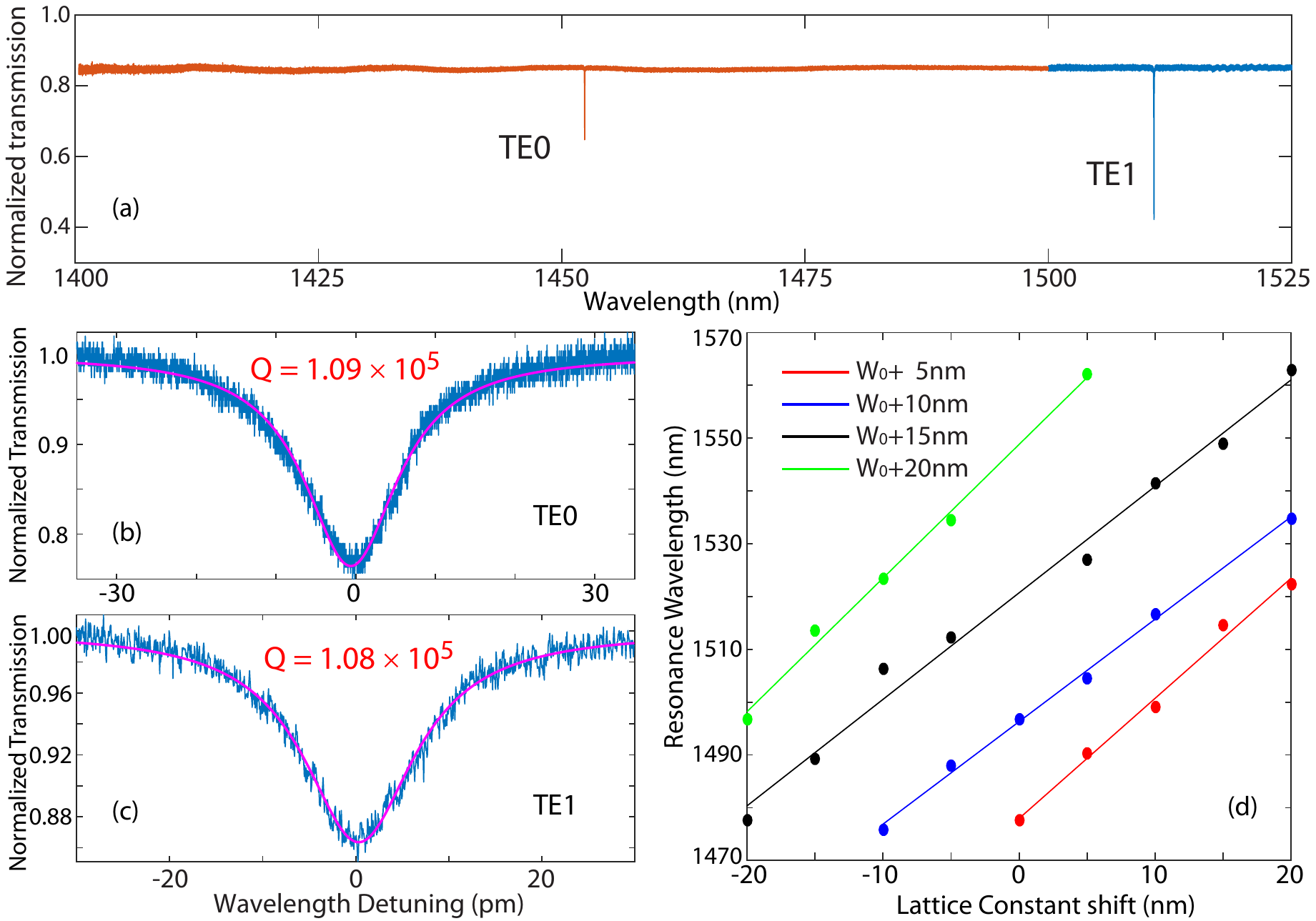}
	\caption{\label{Fig4} Linear optical properties of LN photonic crystal nanocavities. {\bf a.} Laser-scanned transmission spectrum of a LN photonic crystal nanocavity. Two colors on the transmission spectrum indicates the spectral sections scanned by two lasers covering different spectral regions. {\bf b and c.} Detailed transmission spectrum of the fundamental (TE0) and second-order (TE1) cavity modes, respectively, with the experimental data shown in blue and the theoretical fitting shown in red. {\bf d.} Cavity resonance wavelength as a function of lattice constant shift from the nominal values shown in Fig.~\ref{Fig1}(d). Note that, when lattice constant changes, all the lattice constants along the whole nanobeam change by the same amount. Different colors show the cases of different nanobeam widths, varying in a step of 5~nm from the nominal value of ${\rm W_0}$=750~nm. The nanobeam thickness varies accordingly to keep the ratio W/H constant. The dots are experimental data and the solid lines are linear fittings.}
\end{figure}

Figure \ref{Fig3}(a) and (b) show a fabricated device, which clearly show smooth and well defined fine features of the device structure. To characterize the optical property of the device, we launched a continuous-wave tunable laser into the device via evanescent coupling with a tapered optical fiber. Figure \ref{Fig3}(c) shows the schematic of the experimental testing setup, where the optical wave transmitted out from the device is detected by a high-speed detector with a 3-dB bandwidth of 1.3~GHz whose output is characterized by an oscilloscope or an electrical spectrum analyzer, depending on the measured contents. The laser wavelength is calibrated by a Mach-Zehnder interferometer.

By scanning the laser wavelength over a broad telecom band and monitoring the power transmission from the device, we obtained the transmission spectrum of the device shown in Fig.~\ref{Fig4}(a). Figure \ref{Fig4}(a) shows that the device exhibits two high-Q optical resonances at 1452 and 1511~nm, respectively, which correspond to the fundamental and second-order cavity modes (Fig.~\ref{Fig1}(e) and (f)). Detailed characterization of these two modes (Fig.~\ref{Fig4}(b) and (c)) shows that the TE0 and TE1 modes exhibit optical Q as high as $1.09 \times 10^5$ and $1.08 \times 10^5$, respectively. These values are more than two orders of magnitude higher than other LN photonic crystal nanocavities that have ever been reported to date \cite{Baida05, Gu06, Salut06, Gunter09, Pertsch10, Laude10, Baida11, Bernal12, Diziain13, Wang14, Pertsch14}. As discussed in the previous section, the TE0 mode has a radiation-limited optical Q about one order of magnitude higher than the TE1 mode. Therefore, the similarity of optical Qs for these two modes in our devices infers that the optical quality of the devices are still limited by the scattering loss from the sidewall roughness, which can be improved by further optimization of device fabrication.

We are able to precisely control the device dimensions to tune the cavity resonance, as shown in Fig.~\ref{Fig4}(d). On one hand, the cavity resonance depends nearly linearly on the lattice constant. By tuning the lattice constants by an amount between -20~nm and 20~nm in a step of 5~nm from the nominal values shown in Fig.~\ref{Fig1}(d), we are able to shift the cavity resonance wavelength in a linear fashion from 1480~nm to 1560~nm, by a step of about 10~nm (Fig.~\ref{Fig4}(d), black dots). On the other hand, the cavity resonance is sensitive to the width and the thickness of the photonic crystal nanobeam. As shown in Fig.~\ref{Fig4}(d), a similar broadband tuning range of cavity resonance can be obtained by varying simultaneously the width and the thickness of the photonic crystal nanobeam while keeping the ratio of W/H constant.

\section*{Photorefraction and its saturation and quenching}

\begin{figure}[btp]
	\includegraphics[width=0.8\columnwidth]{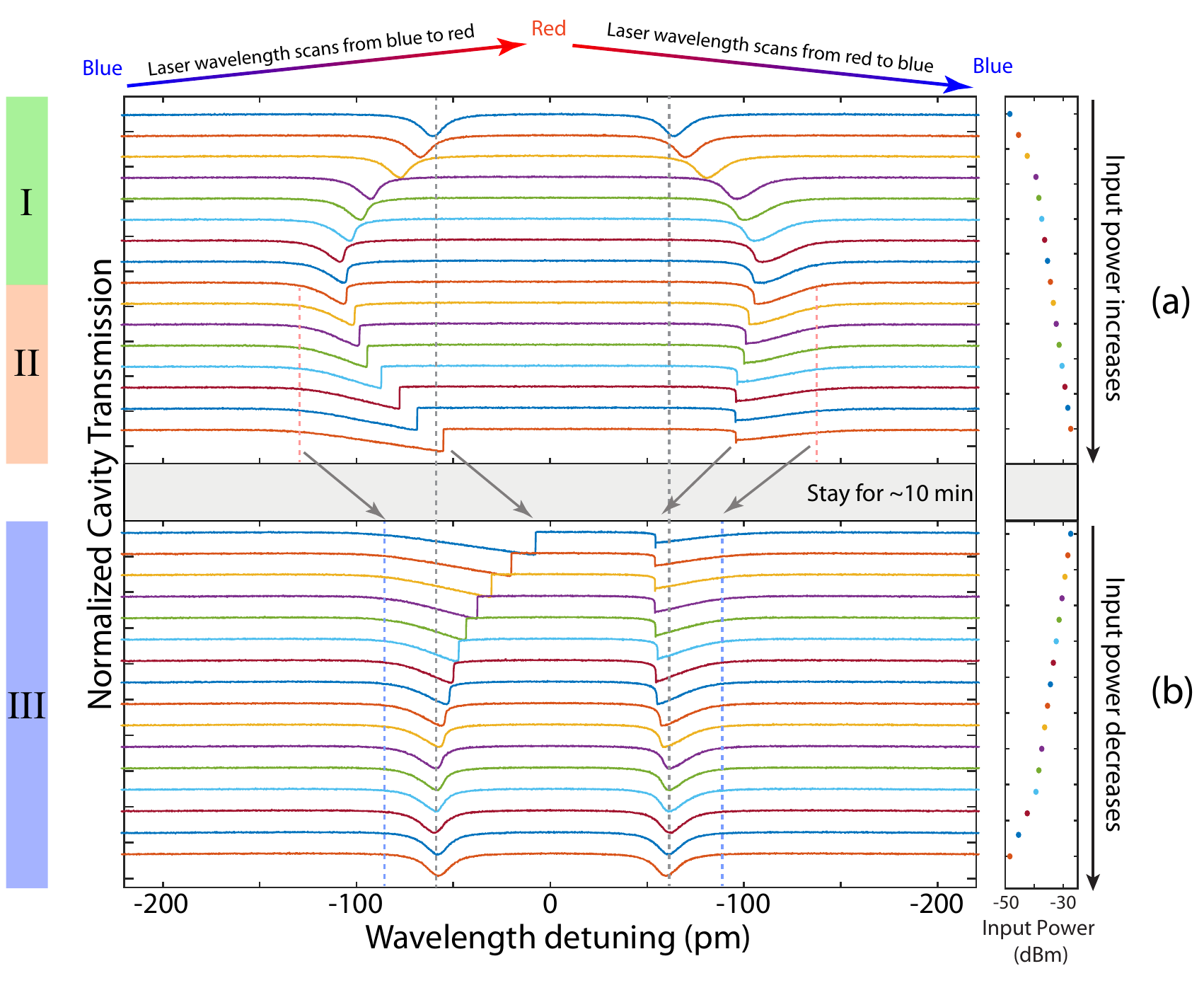}
	\caption{\label{Fig5} Laser-scanned cavity transmission spectrum as a function of input power. The input optical power increases from 330 nW to 41 ${\mu}$W in {\bf a} (from top to bottom), is then maintained at 41 ${\mu}$W for $\sim$10 minutes (gray region), and decreases from 41 ${\mu}$W back to 330 nW in {\bf b} (from top to bottom). The input power corresponding to each scanning spectrum is shown on the right. The laser wavelength is periodically scanned back and forth in a triangular fashion over a spectral range of 280~pm, with a scanning period of 100 ms. The cavity transmission spectra are shifted with each other along the vertical axis for convenient comparison. The color bars on the left illustrate three different regions. In Region I, the optical resonance is blue shifted with increased power, in Region II, the left edge of the optical resonance remains unchanged with increased power, as indicated by the red dashed line. In Region III, the left edge of the optical resonance remains unchanged with decreased power. as indicated by the blue dashed line. During the time period of constant input power (gray region) between Region II and III, the optical resonance red shifts back to its original location, as indicated by the black arrows. The gray dashed line indicates the central location of the optical resonance of the passive cavity in the absence of optical wave. }
\end{figure}

The high quality of the LN photonic crystal nanobeams enables us to observe intriguing nonlinear optical phenomena. Figure \ref{Fig5} shows an example. We scanned the laser wavelength across a cavity resonance back and forth in a periodic triangular fashion, and monitored the transmission of the device. When the input optical power increases from 330 nW to 8 ${\mu}$W, the transmission spectrum changes from a Lorentzian shape to a bistability-type shape while the overall resonance wavelength shifts towards blue by about 55~pm (Fig.~\ref{Fig5}(a), Region I). The bistability-type behavior is simply due to the thermo-optic nonlinearity that responds fairly rapidly to photothermal heating \cite{Vahala04}, which does not affect the overall position of the cavity resonance. The overall blue shift is a typical feature of the photorefractive effect that originates from the electro-optic effect introduced by the space-charge electric field produced via photovoltaic drift current \cite{Gunter06}. The slow relaxation of space charge distribution leads to a net decrease of refractive index which results in an overall blue shift of the cavity resonance \cite{Maleki06, Breunig16, Sun17}.

As the linewidth of the loaded cavity resonance is about 15~pm with a coupling depth of 30 \% while the laser continuously scans over a tuning range of 280~pm, we estimate the average optical power coupled into the cavity is $\sim$133~nW, which corresponds to an averaged energy of $\sim$11.5~aJ and an averaged photon number of only $\sim$87 inside the cavity. This results in a blue tuning rate of $\sim$0.64~GHz/aJ, corresponding to $\sim$84~MHz/photon or $\sim$55~GHz/${\mu W}$, which is 3 orders of magnitude larger than those observed in millimeter-size LN resonators \cite{Maleki06, Breunig16}, clearly showing the dramatically enhanced nonlinear optical effect in LN photonic crystal nanobeam. Such an energy-efficient resonance tuning is of great potential for applications such as all-optical wavelength routing and photonic circuit reconfiguration that are essential for photonic interconnect and optical data communication.

When the input power increases further from 8 ${\mu}$W to 41 ${\mu}$W (Fig.~\ref{Fig5}(a), Region II), although the thermo-optic bistability becomes more profound, as expected, the left edge of the cavity resonance stays at a same wavelength location, as indicated by the red dashed line in Fig.~\ref{Fig5}(a). This infers that the overall cavity resonance wavelength remains unchanged, implying that the photorefraction saturates completely with increased power, in contrast to the photorefraction phenomena observed in other devices  \cite{Maleki06, Breunig16, Sun17}. The underlying mechanism is likely due to the saturation of the generation of space charges responsible for photorefraction, since the extremely tiny physical size of the LN photonic crystal nanocavity leads to a limited number of donors/acceptors that can be excited by optical absorption to produce space charge carriers.

Of particular surprise is that, when we maintained the periodic laser scanning of the cavity mode at an input power of 41 ${\mu}$W, the cavity resonance wavelength moves gradually by itself back to its original value of the passive cavity in the absence of optical power, as indicated by the arrows in Fig.~\ref{Fig5}. After this stage, the overall resonance remains unchanged at its passive value no matter how much optical power is launched into the device, as indicated by the blue dashed line in Fig.~\ref{Fig5}(b) showing the left edge of the cavity resonance. This indicates that the photorefraction is completely \emph{quenched} by the optical wave launched into the device, which has never been observed before. At this state, no matter if we decreased or increased optical power, the phenomena remain same as Fig.~\ref{Fig5}(b), with the overall resonance wavelength nearly intact, except that the extent of thermo-optic bistability varies with optical power. Interestingly, the whole process is reversible. For example, after the photorefraction is quenched, if the device stays at rest for a few hours in the absence of optical wave, it will recover to its original state and all the phenomena shown in Fig.~\ref{Fig5}, such as resonance blue shifting, saturation and quenching of photorefraction, re-appear. The physical nature underlying the observed quenching phenomena is not clear at this moment, which requires further exploration. The quenching of photorefraction would be of great importance for nonlinear optical applications of LN nanophotonic devices, since photorefraction has been shown to be potentially detrimental to nonlinear optical processes \cite{Gunter06, Xu12}.

\section*{Nano-optomechanical properties}

The high quality of the LN photonic crystal nanobeams together with tight optical mode confinement results in strong coupling between the optical field inside the cavity and the mechanical motion of the device structure \cite{Markus14}, which would enable us to probe the optomechanical properties of the device. To do so, we locked the laser wavelength half way into the cavity resonance at the blue detuned side, and monitored the power spectrum of the cavity transmission. The device was tested in the atmospheric environment at room temperature.

\begin{figure}[btp]
	\includegraphics[width=1.0\columnwidth]{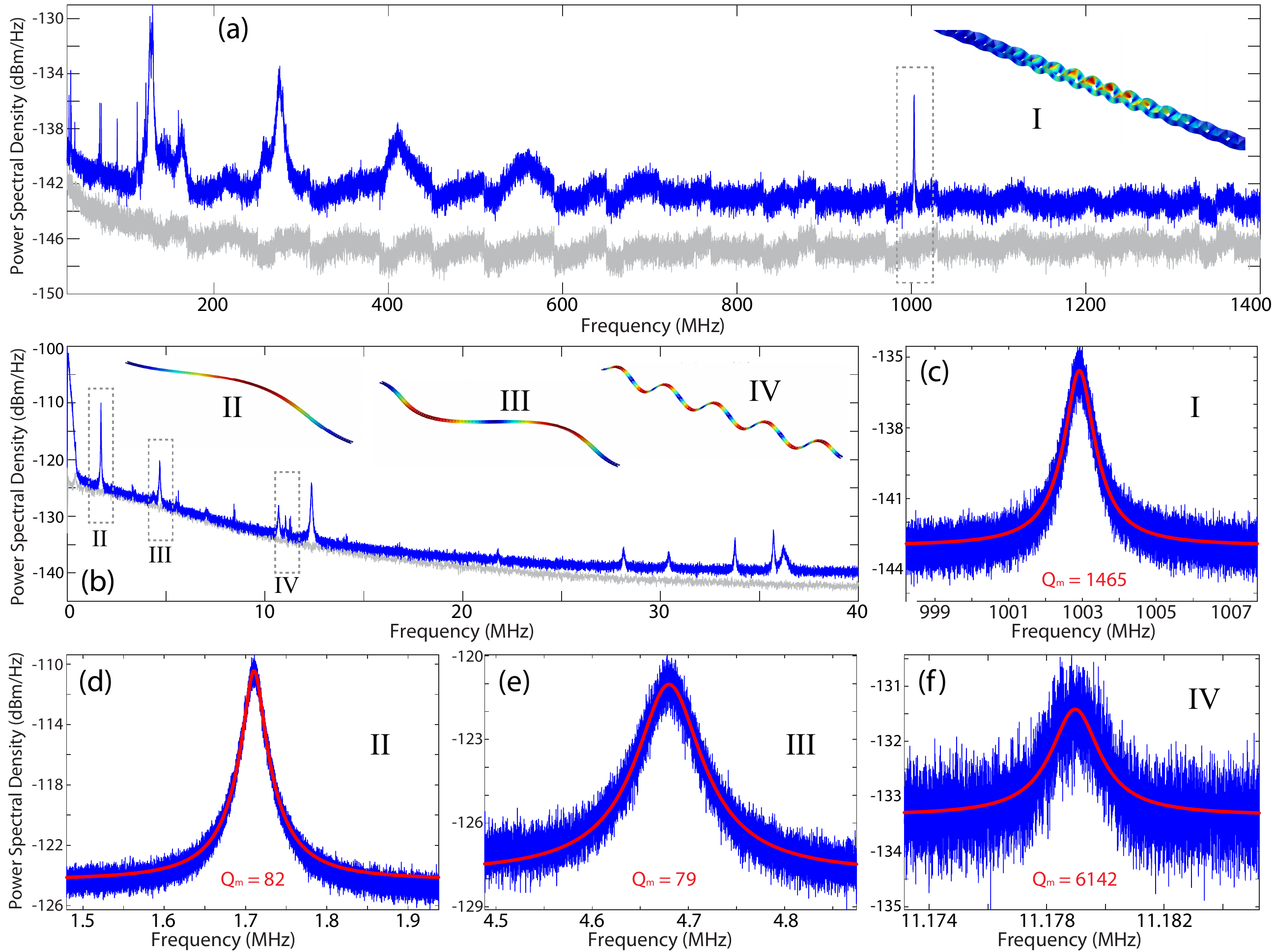}
	\caption{\label{Fig6} Nano-optomechanical properties of a LN photonic crystal nanobeam. {\bf a and b.} Recorded power spectra of cavity transmission over different frequency regions. The insets shows the mechanical displacement profiles of four labeled modes, simulated by the finite-element method. The gray traces show the noise background of the detector. {\bf c-f.} Recorded spectra of mechanical modes at 1.003~GHz, 1.71~MHz, 4.68~MHz, and 11.18~MHz, respectively, with experimental data shown in blue and theoretical fitting shown in red. The mechanical modes are labeled as I--IV in (a) and (b).  }
\end{figure}

Figure \ref{Fig6}(a) and (b) show recorded power spectra of a device, which shows rich mechanical mode families extending over a broad frequency range. As shown in Fig.~\ref{Fig6}(a), the device exhibits a mechanical mode with a frequency at $\frac{\Omega_m}{2\pi} = 1.003$~GHz. Detailed characterization (Fig.~\ref{Fig6}(c)) shows that this mode exhibits an intrinsic mechanical Q of 1465, corresponding to a ${\rm f \cdot Q}$ product of $1.47 \times 10^{12}~{\rm Hz}$, which is comparable to state-of-the-art LN micromechanical resonators \cite{Pijolat09, Piazza13, Jiang16, Bhave15}. We believe that the mechanical damping is dominated by clamping loss, as the device has not been engineered to isolate the mechanical mode from environment. Numerical simulations show that this mechanical mode corresponds to a breathing mode (Fig.~\ref{Fig6}(a), inset) with an effective motion mass of $m_{\rm eff} = 0.81$~picograms and a theoretical frequency of 1.099~GHz. Detailed comparison of the experimental spectrum with theory shows that this mode exhibits an optomechanical coupling coefficient of $\frac{|g_{\rm OM}|}{2\pi} = 22$~GH/nm, which corresponds to a single-photon/single-phonon optomechanical coupling rate of $\frac{|g_o|}{2\pi} = \frac{|g_{\rm OM}|}{2\pi} \sqrt{\frac{\hbar}{2 m_{\rm eff} \Omega_m}} = 71$~kHz. This value is comparable to those observed in most other optomechanical crystals \cite{Painter09, Kartik14, Tang13, Cleland16, Loncar16}, although our devices are not specifically designed for optomechanical applications. It is lower than those in optimized optomechanical crystals reported in \cite{Painter12, Kartik16} that were optimized to enhance the photoelastic contribution. As LN exhibits outstanding acousto-optic property \cite{Gaylord85}, we expect that future optimization of device design would be able to significantly improve the optomechanical properties of the LN photonic crystal nanobeams.

On the other hand, detailed characterization of low-frequency modes (Fig.~\ref{Fig6}(b)) shows that a majority of them exhibit low mechanical qualities in the order of $\sim100$, which is primarily due to air damping since low-frequency mechanical modes exhibit large amplitudes of thermal mechanical motion, sensitive to air damping. Two examples are given in Fig.~\ref{Fig6}(d) and (e), where the modes at 1.71~MHz and 4.68~MHz exhibits mechanical Qs $\sim$80. Numerical simulations show that these two modes correspond to the first-order and second-order flexural modes (Fig.~\ref{Fig6}(b), inset I and II), respectively, with effective motional masses of 7.2 and 7.9~picograms. Comparison of the experimental spectra with theory shows that these two modes exhibit $\frac{|g_{\rm OM}|}{2\pi}=$ 0.35 and 0.45~GHz/nm, respectively, corresponding to $\frac{|g_o|}{2\pi}=$ 9.1 and 6.8~kHz. The small values of optomechanical coupling are primarily due to the nature of the mechanical modes (Fig.~\ref{Fig6}(b), inset I and II) which do not couple well with the optical cavity mode localized at the beam center. Figure~\ref{Fig6}(f) shows that a mechanical mode at 11.18~MHz shows a high mechanical Q of 6142, which is likely to be a high-order flexural mode (Fig.~\ref{Fig6}(b), inset III) that is not as sensitive to air damping as other modes.

\section*{Conclusion and discussion}

In summary, we have demonstrated LN photonic crystal nanobeam resonators with optical Q up to $10^5$ that is more than two orders of magnitude higher than other LN photonic crystal nanocavities reported to date \cite{Baida05, Gu06, Salut06, Gunter09, Pertsch10, Laude10, Baida11, Bernal12, Diziain13, Wang14, Pertsch14}. The devices exhibit an effective mode volume as small as $\sim 1.03(\lambda/n)^3$. The high optical Q together with tight optical mode confinement results in intriguing nonlinear optical phenomena. We have observed significant cavity resonance tuning induced by the photorefractive effect, with a tuning rate of $\sim$0.64~GHz/aJ, corresponding to $\sim$84~MHz/photon, three orders of magnitude greater than other LN resonators \cite{Maleki06, Breunig16}. In particular, the devices exhibit strong saturation and quenching of photorefraction that has never been observed before. Photorefraction-induced optical damage is known to be detrimental to nonlinear optical processes in LN crystals \cite{Gunter06, Xu12}, which has become a major obstacle to LN nonlinear photonics. Conventional approaches to mitigate photorefraction is to dope LN crystal with certain ions to increase the photorefraction threshold \cite{Xu12}. The strong saturation and quenching of photorefraction observed in our devices might offer an elegant solution to this problem, making LN nanophotonic devices particularly promising for nonlinear photonic applications.

On the other hand, the demonstrated devices exhibit strong coupling between the optical cavity mode and the mechanical motion of the device structures, with which we were able to characterize the rich nanomechanical motions of the device. We observed mechanical modes with frequency up to 1.003~GHz with a ${\rm f \cdot Q}$ product of $1.47 \times 10^{12}~{\rm Hz}$ that is comparable to state-of-the-art LN micromechanical devices \cite{Pijolat09, Piazza13, Jiang16, Bhave15}. The devices exhibit a single-photon/single-phonon optomechanical coupling rate of $\frac{|g_o|}{2\pi} = 71$~kHz that is comparable to most other optomechanical crystals \cite{Painter09, Kartik14, Tang13, Cleland16, Loncar16}, although our devices are not specifically designed for optomechanical applications. LN exhibits strong piezoelectric effect, electro-optic effect, and electromechanical coupling, significantly larger than other materials such as aluminum nitride and gallium arsenide \cite{Gaylord85, Piazza13, Mateti09}. Therefore, LN photonic crystals would offer a promising device platform that could achieve mutual strong couplings between electrical, optical, and mechanical degrees of freedom for various optoelectronic, optomechanical, and electromechanical applications.

\section*{Funding Information}
National Science Foundation (ECCS-1509749); Defense Advanced Research Projects Agency SCOUT program (W31P4Q-15-1-0007); the Open Program (Grant No. 2016GZKF0JT001) of the State Key Laboratory of Advanced Optical Communication Systems and Networks at Shanghai Jiao Tong University, China.

\section*{Acknowledgments}

This study was performed in part at the Cornell NanoScale Science and Technology Facility (CNF), a member of the National Nanotechnology Infrastructure Network.

\begin{footnotesize}

\end{footnotesize}

\end{document}